\documentstyle[12pt]{article}
\input{psfig}

\newcommand {\beq} {\begin{equation}}
\newcommand {\eeq} {\end{equation}}

\title{Using the Uncharged Kerr Black Hole as a Gravitational
 Mirror }    

\author{{Claes R Cramer\thanks{e-mail C.R.Cramer@qmw.ac.uk}}\\
\small{\em Department of Theoretical Physics,}\\{Lund University}
,\\ \small{\em S\"olvegatan 14A}\\
\small{\em SE-223 62 Lund, Sweden}\\
\small{\em and}\\
\small{\em School of Mathematical Sciences,}\\{\em Queen Mary 
\& Westfield
College},\\ \small{\em Mile End Road}\\
\small{\em London E1 4NS, UK}}

\begin{document}

%\sloppy

\maketitle

\begin{abstract}
We extend the study of the possibility to use the Schwarzschild
black hole as a gravitational mirror to the more general case 
of an uncharged Kerr black hole. We use the null geodesic equation 
in the equatorial plane to prove a theorem concerning
the conditions the impact parameter has to satisfy if there 
shall exists boomerang photons. 
We derive an equation for these boomerang photons and an equation 
for the emission angle. Finally, the radial null geodesic equation 
is integrated numerically in order to illustrate boomerang photons. 

\end{abstract}

\newpage
\section{Introduction}
The  gravitational field surrounding a black hole can be 
so severe that photons entering the field will circulate around 
the black hole once or several times before they either escape 
to infinity or pass the event horizon. This suggests the possibility 
of using a black hole as a gravitational mirror in the sense that 
images are produced when photons return to their emission point. 
These gravitational mirror effects produced by a Schwarzschild 
black hole have been studied by W. M. Stuckey \cite{Stuckey93}. 
The present paper extends Stuckey's study in the sense that 
we investigate the possibility of using the uncharged Kerr
black hole as a gravitational mirror.\\\\
In section 2 we briefly review the Kerr solution. 
In section 3 we formulate an existence condition for
boomerang photons and use this condition in order to prove a theorem
related to  the corresponding conditions for the impact parameter. 
Furthermore, we derive an integral equation for boomerang photons 
and an equation for the emission angle. In section 4 we integrate 
the radial null geodesic equation in order to illustrate boomerang 
photons.

\section{The Kerr Solution}
The Kerr solution is a well-known stationary and axial symmetric solution 
of Einstein's vacuum field equations. The solution can be taken to 
represent a rotating charged black hole and the line element, in 
the case of vanishing charge and with Boyer-Lindquist coordinates, 
takes the form
\beq
ds^{2}=\frac{1}{\Sigma}\Delta(dt-asin^{2}\theta d\phi)^{2}-
\Sigma(\frac{1}{\Delta}dr^{2}+d\theta^{2})-
\frac{1}{\Sigma}sin^{2}\theta \left ((r^{2}+a^{2})
d\phi-adt \right)^{2} \label{line}
\eeq
where
\begin{eqnarray}
\Sigma=r^{2}+a^{2}cos^{2}\theta,\\
\Delta=r^{2}-2Mr+a^{2}.
\end{eqnarray}
The parameter $M$ is the geometrical mass of the black hole, while 
$a$ is a parameter defined by
\beq
a:=\frac{J}{M}
\eeq
where $J$ is the total angular momentum of the black hole. It will be
assumed throughout the paper that $a\not=0$ since when $a$ equals zero
we recover the Schwarzschild case already treated by W. M. Stuckey  
\ \cite{Stuckey93}.\\\\
A calculation of the Riemann scalar shows that $\Sigma=0$ represents a ring 
singularity while $\Delta=0$ represents singularities arising from the
choise of coordinates, see e.g.\ \cite{Chandrasekhar83}.
The equation
\beq
\Delta=0
\eeq
has the real roots
\beq
r_{\pm}=M\pm\sqrt{m^{2}-a^{2}}
\eeq
where $r_{+}$ represents the event horizon and $r_{-}$ the Cauchy horizon, 
see e.g. Chandrasekhar for details \ \cite{Chandrasekhar83}.\\\\

We shall need the null geodesic equation in the equatorial plane, 
i.e. $\theta=\pi/2$, in our study so we might as well
derive it even though the derivation can be found in textbooks. 
The easiest way to derive the equation is to use a proposition
given in Wald  which states that the inner product between a 
Killing vector and a tangent to a geodesic is constant along the 
geodesic \ \cite{Wald84}. We associate with this constant  a 
conserved physical quantity. A quick look at the line element 
immediately reveals that we can take
\beq
\delta^{a}_{t}=(1,0,0,0)
\eeq
and
\beq
\delta^{a}_{\phi}=(0,0,0,1)
\eeq 
as Killing vectors representing energy $E$ and angular momentum $L$. 
This follows since the metric components are independent of $t$ and 
$\phi$ and hence the metric field remains unchanged along the 
integral curves of the 
given vectors. Thus for 
\beq
u^{a}=\frac{dx^{a}}{d\tau}
\eeq
where
\beq
x^{a}=(t,r,\theta,\phi)
\eeq
we easily find that the
constants of motion are given by
\begin{eqnarray}
E=(1-\frac{2M}{r})\dot{t}+\frac{2Ma}{r}\dot{\phi} \label{energy}\\
L=(a^{2}+r^{2}+\frac{2Ma^{2}}{r})\dot{\phi}-\frac{2Ma}{r}\dot{t} 
\label{angmom}
\end{eqnarray} 
where ($\tau$ being a parameter along the photon world line)
\begin{eqnarray}
\dot{t}=\frac{dt}{d\tau}\\
\dot{\phi}=\frac{d\phi}{d\tau}.
\end{eqnarray} 
Using equations (\ref{energy}) and (\ref{angmom}) and the null 
geodesic equation
\beq
g_{ab}u^{a}u^{b}=0
\eeq
one finds the following equation for radial null geodesics in 
the equatorial plane
\beq
\dot{r^{2}}=E^{2}+\frac{2M}{r^{3}}(L-aE)^{2}-\frac{1}
{r^{2}}(L^{2}-a^{2}E^{2}). \label{rad}
\eeq
\section{Boomerang photons}
If an uncharged rotating black hole is to cause mirror images in 
the equatorial plane there has to exist a possibility for 
emitted photons to return to their emission point. 
We shall call such photons {\em boomerang photons} or to be more
precise we define a $n$-fold boomerang photon as follows.\\\\
{\bf Definition}\\ A photon in the equatorial plane, circulating outside the
event horizon $r_{+}$ $n$ times, where $n$ is a positive integer, before
it returns to its emission point $(r_{0},\phi_{0})$ is called an $n$-fold 
boomerang photon.\\\\ 
Using this definition we immediately notice that the following condition 
has to be satisfied if boomerang photons are to exist.\\\\
{\bf Condition}\\ A necessary condition for the existence of an $n$-fold 
boomerang photon satisfying the initial condition
\beq
|\dot{r_{0}}|\neq 0
\eeq
is that $\dot{r}$ changes sign for some $r\in[r_{+},\infty[$.
Given this existence condition we now prove the following theorem.\\\\
{\bf Theorem}\\
If turning points where $\dot{r}$ changes sign exist then the impact 
parameter $D$ defined by
\beq
D:=\frac{L}{E}
\eeq
has to satisfy 
\beq
D>6Mcos\frac{\varphi}{3}-a \label{ineqone}
\eeq
or
\beq
D<6Mcos(\frac{\varphi}{3}+\frac{2\pi}{3})-a \label{ineqtwo}
\eeq
where
\beq
cos\varphi=-\frac{a}{M}
\eeq
and
\beq
0<a^{2}\leq M^{2}.
\eeq
{\bf Proof}\\
If $\dot{r}$ changes sign then it follows from the radial null 
geodesic equation (\ref{rad}) that the equation
\beq
f(r;D;a)\equiv r^{3} -(D^{2}-a^{2})r +2M(D-a)^{2}=0 \label{humpi}
\eeq
has to have solutions. We shall only consider real positive roots as being of
physical importance. We shall also exclude the case of multiple roots since 
that corresponds to non-stable circular curves, as will be explained
in section 4.\\\\
Using Descartes' rule of signs which states that the number of real positive 
roots of a polynom with real coefficients is never greater than the number 
of sign variations in the sequence of its coeffiencents and if less 
always by an even number, see e.g. Uspensky \ \cite{Upensky48} for 
a proof of this statement,  we conclude that equation (\ref{humpi})  
has none or two real roots if and only if

\beq
D^{2}>a^{2}.
\eeq
Now, there will be two positive real roots if the discriminant of equation
(\ref{humpi}), i.e. 
\beq
d=-\frac{(y-2a)^{3}}{27}(y-6Mcos\frac{\varphi}{3})(y-6Mcos
(\frac{\varphi}{3}+\frac{2\pi}{3}))(y-6Mcos(\frac{\varphi}{3}+\frac{4\pi}{3}))
\eeq
satisfies the {\em casus irreducibilis} inequality 
\beq
d<0
\eeq
and if $D^{2}>a^{2}.$\\\\
Here we have introduced the symbols
\beq
y:=D+a
\eeq
and
\beq
cos\varphi:=-\frac{a}{M}.
\eeq
This is true since firstly,
\beq
f(0;D;a)>0
\eeq
and secondly, when
\beq
r \rightarrow -\infty
\eeq
then
\beq
f \rightarrow -\infty
\eeq
monotonely for $r<r_{max}=-\sqrt{\frac{1}{3}(D^{2}-a^{2})}$. 
Finally, using the  casus irreducibilis inequality we immediately 
conclude that the impact parameter is subject to  inequality 
(\ref{ineqone}) or (\ref{ineqtwo}) $\Box$.\\\\ 
We have so far shown that a necessary condition for the existence of
 boomerang photons is that
inequality (\ref{ineqone}) or (\ref{ineqtwo}) is satisfied.
We shall now proceed to derive an equation for boomerang photons.
\\\\ Assume that an $n$-fold boomerang photon $\gamma$ is emitted 
at $(r_{0},0)$ in the equatorial plane and that the impact parameter 
for this photon is $D_{n}$. Using equations (\ref{energy}),
(\ref{angmom}), (\ref{rad})
and 
\beq
\dot{r}=\frac{dr}{d\phi}\dot{\phi}
\eeq
then we get
\beq
\frac{d\phi}{dr}=\pm\frac{\frac{2Ma}{r}+(1-\frac{2M}{r})D_{n}}
{(r^{2}-2Mr+a^{2})(1+\frac{2M}{r^{3}}(D_{n}-a)^{2}-
\frac{1}{r^{2}}(D^{2}_{n}-a^{2}))^{\frac{1}{2}}} \label{dumpi}
\eeq
where the negative sign is taken to  represent ingoing photons and the 
plus sign outgoing photons.
If $\gamma$ is emitted towards the black hole and
if $r_{s}$ is a turning point then using equation (\ref{dumpi}) we find
\beq
\phi(r_{s})=-\int^{r_{s}}_{r_{0}}dr\frac{\frac{2Ma}{r}+
(1-\frac{2M}{r})D_{n}}
{(r^{2}-2Mr+a^{2})(1+\frac{2M}{r^{3}}(D_{n}-a)^{2}-
\frac{1}{r^{2}}(D^{2}_{n}-a^{2}))^{\frac{1}{2}}}. \label{da1}
\eeq
Furthermore, if $\gamma$ returns to its emission point then from equation 
(\ref{dumpi})
\beq
2\pi n-\phi(r_{s})=\int^{r_{0}}_{r_{s}}dr\frac{\frac{2Ma}{r}+
(1-\frac{2M}{r})D_{n}}
{(r^{2}-2Mr+a^{2})(1+\frac{2M}{r^{3}}(D_{n}-a)^{2}-
\frac{1}{r^{2}}(D^{2}_{n}-a^{2}))^{\frac{1}{2}}} \label{da2}
\eeq
where
\beq
n=1,2,3,...
\eeq
Hence, using equations (\ref{da1}), (\ref{da2}) we get the following 
elliptic integral for an $n$-fold boomerang photon
\beq
2\pi n=\mp \int^{r_{s}}_{r_{0}}dr\frac{\frac{2Ma}{r}+(1-\frac{2M}{r})D_{n}}
{(r^{2}-2Mr+a^{2})(1+\frac{2M}{r^{3}}(D_{n}-a)^{2}-
\frac{1}{r^{2}}(D^{2}_{n}-a^{2}))^{\frac{1}{2}}}
\eeq
where the plus sign follows trivially from an analogous discussion 
where $\gamma$ is emitted outwards.\\\\
Notice that an $n$-fold boomerang photon changes direction, i.e. 
from inwards to outwards or conversely, when $\phi=n\pi$. 
One could solve the above elliptic integral in order to investigate
whether or not the inequalities (\ref{ineqone}) and (\ref{ineqtwo}) 
are sufficient conditions for the existence of boomerang photons. 
We shall not pursue this question in this paper.
Instead we prefer a numerical study in order to illustrate boomerang photons.
However, before we consider a numerical illustration of boomerang photons we 
give an equation for the emission angle. An observer who is emitting photons 
will measure an angle, say $\delta$ between the photon trajectory and the 
radial direction. The observer is assumed to be at rest at $(r_{0},\phi_{0})$ 
in the equatorial plane. This means that 
\beq
d\tau^{2}=\frac{1}{r^{2}}(\Delta-a^{2})dt^{2}.  
\eeq
If we now use  the null geodesic equation and the equation above then 
we get after some algebra that the emission angle is given by
\beq
\delta=atan{\left ( \frac{\Delta^{\frac{1}{2}}}{r^{2}}
((r^{2}+a^{2})^{2}-a^{2}\Delta)^{\frac{1}{2}}\frac{d\phi}{dr} \right )}.
\eeq

\section{Numerical considerations}
In a Schwarzschild field we can use a black hole to produce ring-shaped mirror 
images \ \cite{Stuckey93}. This is a direct consequence of the spherical 
symmetry property of a Schwarzschild field. If we consider null geodesics 
in an axial symmetric space-time such as the Kerr space-time
then we cannot sure of that the geodesics generally remain in a plane. 
Hence, even if we show that boomerang photons
exist in the equatorial plane, we cannot conclude that such photons exist 
in other planes. Furthermore, even if they exist then we will not be able 
to produce ring-shaped images since that requires
spherical symmetry. In fact, we can conclude that
due to the rotation of  the Kerr
black hole,  point images produced by clockwise boomerang photons and 
counter-clockwise boomerang photons will not be placed symmetrically
with respect to the symmetry axis, see figure 1. 
\begin{figure}[b]
      \hbox{ \vbox{
        \begin{center}
         \mbox{ \psfig{figure=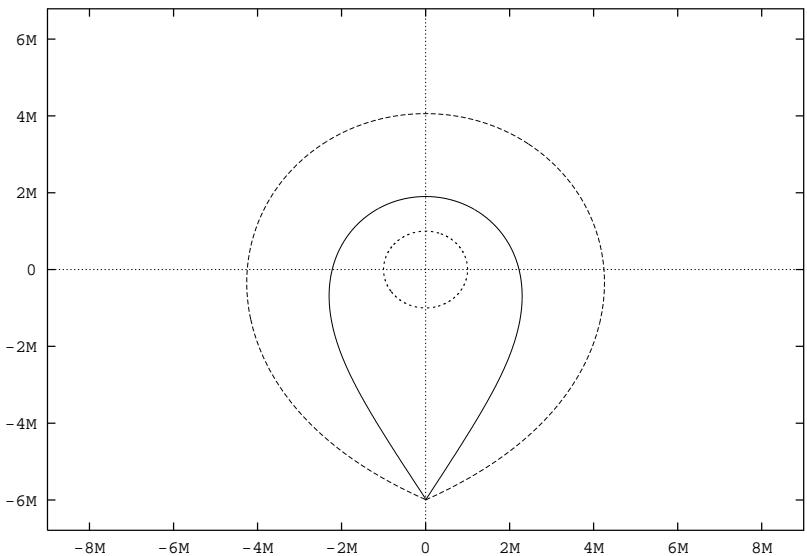,width=8.5cm,height=8.5cm}}
         \end{center}
          }
          }
\caption{\em Boomerang photons in an extreme Kerr field. These photons
are emitted from $6M$. The photon illustrated as a dashed curve
orbits the black hole clockwise. The other photon orbits the
black hole counter-clockwise. The inner, dashed circle represents
the event horizon and the rotation direction of the black hole is
counter-clockwise.}
\end{figure}
It might come as a surprise that there is no distorsion of the symmetry,
i.e. broken reflection symmetry, in the photon orbit. 
The reason for this is that the photon
will feel the same 'gravitational vind' effect whether or not the 
coordinate position is $(r, \phi)$ or $(r,-\phi)$. This property
can also be seen from equations (\ref{da1}), (\ref{da2}) since the
right hand side of these two equations are the same. 
\newpage
If the impact parameter equals
\beq
D=6Mcos\frac{\varphi}{3}-a
\eeq
or
\beq
D=6M\cos{\left ( \frac{\varphi}{3}+\frac{2\pi}{3} \right )}
\eeq
then it follows that the real and positive roots of $f$ are equal. 
We can give this case a physical meaning, if we notice that the value 
of a double root corresponds to a local minimum of $f$. Thus, if a photon 
is emitted at this point with an impact parameter given by one of the 
equations above it will orbit the black hole in a circular orbit. 
This orbit we call a {\em photon circle}. 
The photon circle is non-stable since a local minimum of $f$ corresponds 
to a local maximum in an effective potential. Furthermore, a photon orbiting 
$n$-times in a photon circle is by definition a $n$-fold boomerang photon,
but such a photon has to be emitted at a right angle with respect to the 
radial direction.
Now, if an observer that is emitting boomerang photons 
moves towards the photon circle from infinity he or she will notice as in the 
Schwarschild case that the emission angle increases and approaches $\pi/2$
as he reaches the photon circle. Thus, the point images will be spread 
out as the observer moves towards the photon circle. If the observer 
continues towards the horizon, the point images
appear in the direction opposite the black hole, unless the photon 
circle coincides with the event horizon as it does in an extreme Kerr 
field where $a$ equals $M$, see e.g. figure 2 and table 1.
\newpage
\begin{figure}[t,h]
      \hbox{ \vbox{
        \begin{center}
         \mbox{ \psfig{figure=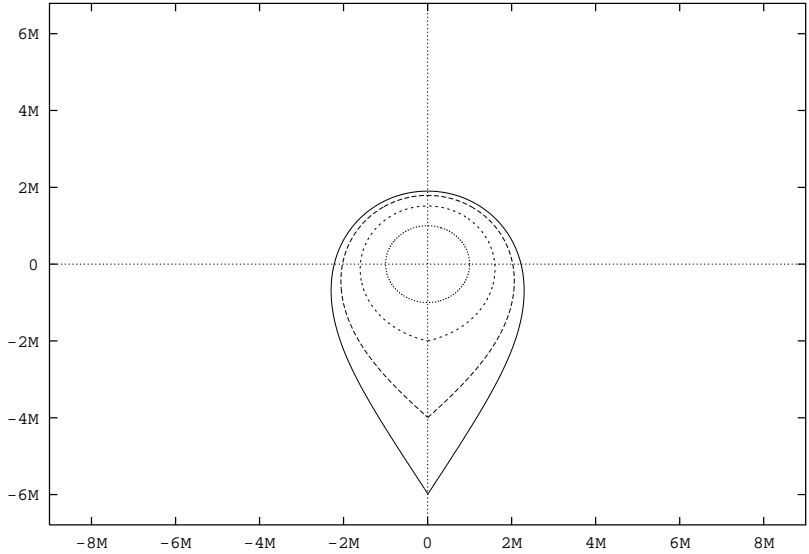,width=8.5cm,height=8.5cm}}
         \end{center}
          }
          }             
\caption{\em Boomerang photons emitted counter-clockwise in an extreme
Kerr field. Notice how the emission angle increases as the emitter moves
towards the event horizon, which in this case coincides with the photon circle
at $M$}
\end{figure}

\ \ \ \ \ \ \ \ \ \ \ \ \ \ \ \ \ \ \ \ \ 
\ \ \ \ \ \ \ \ \ \ \begin{tabular}{||c|c|c|c||}\hline
$r/M$&$|r-r_{num}|/M$&$\delta$&
$D/M$\\ \hline
2&$1.252\cdot 10^{-10}$	&68.05&2.513\\ \hline
4&$5.895\cdot 10^{-9}$&$36.76$&2.788\\ \hline
$6$&$2.513\cdot 10^{-8}$&$26.91$&2.900\\ \hline  
\end{tabular}\\\\
{\em Table 1. Numerical results for boomerang photons in an 
extreme Kerr field.
Notice that from equation (40) the critical value of $D$ is $2M$, 
and that the numerical 
results for the impact parameter satisfy $D>2M$.}
\newpage
In all previous figures we have only considered boomerang photons 
orbiting the black hole once, i.e the case when the turning angle 
equals $\pi$. We shall now illustrate boomerang photons
orbiting the black hole twice before returning to their emission point. 
From the elliptic integral it follows that a photon orbiting a black hole 
twice will cross its own curve at
$\phi=\pi$ and that the turning angle equals $2\pi$, see figure 3.
\begin{figure}[t,h,b]
      \hbox{ \vbox{
        \begin{center}
         \mbox{ \psfig{figure=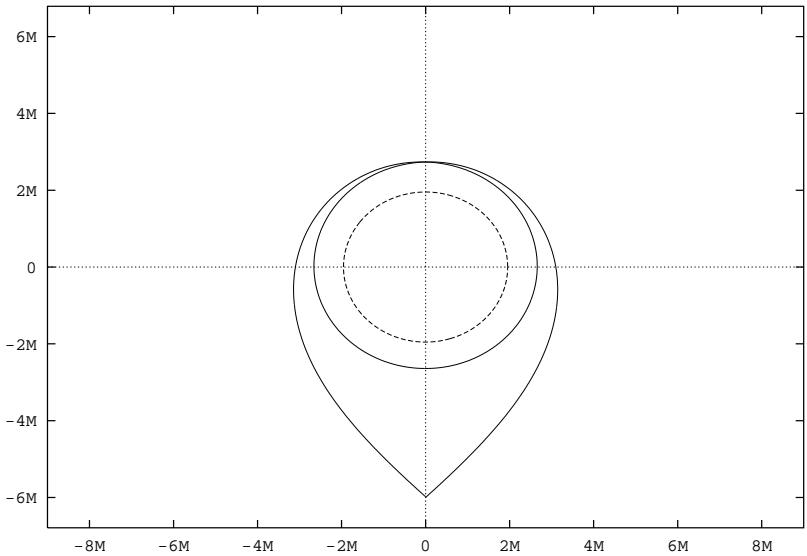,width=8.5cm,height=8.5cm}}
         \end{center}
          }
          }
\caption{\em A Boomerang photon orbiting a Kerr black hole twice. The
rotation parameter $a$ equals $0.3M$.}
\end{figure}

The numerical results were obtained using the Runge-Kutta quartic 
method \ \cite{Froberg85}.
The ordinary differential equation solved was
\beq
\frac{dr}{d\phi}=\pm \frac{\Delta \left 
(1+\frac{2M}{r^{3}}(D-a)^{2}-\frac{1}{r^{2}}(D^{2}-a^{2})
\right )^{1/2}}{\frac{2Ma}{r}+(1-\frac{2M}{r})D}
\eeq
where the sign was reversed when
\beq
\dot{r}=0
\eeq
{\bf Acknowledgement}\\
I have benefited from numerous interactions with my thesis advisor 
Professor B. E. Y. Svensson while writing this work. Therefore, 
I would like to express my gratitude to him. I also wish to thank 
Professor M. A. H. MacCallum for making comments on this revised version 
of my master thesis work. Finally, I would like to thank research student 
Jari H\"akkinen for helping me with all the subtle problems I
have had with the computer.

\end{document}